# Robust parameter design for Wiener-based binaural noise reduction methods in hearing aids

Diego M. Carmo, Ricardo Borsoi, and Márcio H. Costa, *Member, IEEE*

*Abstract*—This work presents a method for designing the weighting parameter required by Wiener-based binaural noise reduction methods. This parameter establishes the desired trade-off between noise reduction and binaural cue preservation in hearing aid applications. The proposed strategy was specially derived for the preservation of interaural level difference, interaural time difference and interaural coherence binaural cues. It is defined as a function of the average input noise power at the microphones, providing robustness against the influence of joint changes in noise and speech power (Lombard effect), as well as to signal to noise ratio (SNR) variations. A theoretical framework, based on the mathematical definition of the homogeneity degree, is presented and applied to a generic augmented Wiener-based cost function. The theoretical insights obtained are supported by computational simulations and psychoacoustic experiments using the multichannel Wiener filter with interaural transfer function preservation technique (MWF-ITF), as a case study. Statistical analysis indicates that the proposed dynamic structure for the weighting parameter and the design method of its fixed part provide significant robustness against changes in the original binaural cues of both speech and residual noise, at the cost of a small decrease in the noise reduction performance, as compared to the use of a purely fixed weighting parameter.

## I. INTRODUCTION

HEARING aids are electronic prosthesis which compensate mild to moderate hearing losses [1]. According to [2], more than 80 percent of hearing-impaired people have both ears affected by a reduction in hearing ability, requiring the simultaneous use of two hearing aid devices. For these people, the best results in hearing loss compensation are achieved through the use of binaural hearing aids, which allow for the exchange of data and control parameters between the left and right gadgets.

Binaural noise reduction methods consist of recovering the signal of interest (i.e., speech) from multiple noisy observations at both ears [3]. Such approach does not only lead to increased noise reduction, as compared to the bilateral use of hearing aids, but also allows for the preservation of the original localization cues of the sound sources, resulting in reliable spatial perception [4]. For spatially separated target and undesired sources, the user may be favored by the spatial-release-from-masking effect [5], leading to improved speech intelligibility [6] as compared to situations in which these sources are near each other.

Binaural noise reduction methods can be roughly classified as spectral postfiltering or spatial filtering approaches. The postfiltering approach consists of techniques that, directly or indirectly, apply the same spectral-temporal gain to all noisy-input signals at each bin or frequency band [7] [8] [9] [10]. This leads to a perfect preservation of the instantaneous auditory information [4]. However, the noise reduction results obtained by such techniques are similar to those provided by bilateral systems [4]. Spatial filtering techniques, on the other hand, may apply distinct filters to the input signals, which can significantly increase speech quality and intelligibility. However, this comes at the cost of some distortion in the perception of the acoustic scenario.

The binaural multichannel Wiener filter (MWF) is the most studied approach for binaural noise reduction. It has been demonstrated that for an acoustic scenario comprised of one speech and one noise acoustic point sources, the MWF significantly reduces the power of the undesired signal, while naturally preserving the binaural cues of the target speech. However, as a side-effect, the MWF distorts the localization cues of the noise source, changing its perceived localization with relation to the azimuth angle of the speech source [11] [12] [13].

Three major binaural cues are employed by the human auditory system for the localization of acoustic sources. They are the interaural time difference (ITD), the interaural level difference (ILD) and the interaural coherence (IC). The ITD quantifies differences in the time of arrival of the sound waves as they reach both ears [14]. It is considered as the main localization cue at low frequencies ($< 1500\,\text{Hz}$), since wavelengths within this range are longer than the average size of the human head. The ILD measures differences (in logarithmic scale) between sound intensities in the two ears [14]. It is more relevant at high frequencies ($> 1500\,\text{Hz}$), in which the amplitude of the sound waves is significantly affected by reflection, diffraction, and scattering from the human head and torso. The IC is defined as the normalized cross-correlation between signals at both ears, and determines the reliability of

This work was supported by the Brazilian Ministry of Science and Technology (CNPq) under grant 315020/2018-0.

D. M. Carmo and R. Borsoi are with the Graduate Program in Electrical Engineering, Federal University of Santa Catarina, Florianopolis-SC, 88040-370, Brazil (e-mails: diego.carmoh@gmail.com, raborsoi@gmail.com). M. H. Costa is with the Department of Electronic and Electrical Engineering, Federal University of Santa Catarina, Florianopolis-SC, 88040-370, Brazil (e-mail: costa@eel.ufsc.br).

the ILD and ITD cues in multisource and reverberant scenarios [15]. For diffuse acoustic signals, the IC is the prevalent binaural cue [12].

A large and widely employed subclass of MWF-based binaural noise-reduction methods aim to preserve both speech and noise localization cues. It is based on spatial filters calculated as the solution to an optimization problem which minimizes the weighted sum of two specific cost functions. The first one is the MWF cost function, which aims to promote noise reduction at a limited speech distortion. The second one is a penalization term related to binaural cue preservation, which aims to retain the auditory impression of the residual interfering noise in the processed signal. The relative weight between both terms of this augmented cost function is determined by an adjustable parameter, whose setpoint depends on the statistical characteristics of the received signals and on the acoustic scenario. This parameter must be carefully selected to establish the desired trade-off between both the noise reduction and binaural cue preservation objectives.

In [16], an augmented MWF-based cost function was proposed considering two additive penalty terms to preserve both the ITD and the ILD of the processed noise, which are known to be the prevalent binaural cues for localization of point sound sources in the median plane of the head. In [17], the preservation of both the ITD and ILD was obtained using a single penalty term, which considered the interaural transfer function (ITF). The ITF carries information from both ITD and ILD binaural cues [18]. However, this method was shown to be incapable of preserving the spatiality of diffuse noise fields. To address this issue, the IC cost function was proposed in [12] for the preservation of environmental diffuse noise. In [19], it was shown that the IC cost function can also be used to preserve the ITD of point sources with improved lateralization performance in comparison to the approach presented in [16].

Solutions obtained by augmented MWF cost functions with fixed weight parameters present adequate performance in acoustic scenarios, comprised of one speech and one noise source under stationary and time-invariant conditions. However, they show significant performance degradation (of the desired trade-off between noise reduction and spatial preservation of the noise source) under nonstationary situations. Such conditions are intrinsic of spoken communication, in which unpredictable changes of the environmental noise may lead to acoustic power variations, sometimes compelling the speaker to adjust the speech level to maintain intelligible communication [20]. This phenomenon is known as *Lombard effect* [21] [22] [23], and affects a number of speech features, such as: voice intensity; spectral slope of glottal waveforms; formant locations and bandwidths; and energy ratios in voiced/unvoiced phonemes [24] [25]. Although this is a well-studied problem in the speech recognition area, it has not received enough attention in the binaural noise reduction field.

A review on MWF-based binaural techniques shows that most binaural-cue penalty terms presented in the literature are invariant to input power variations [12] [16] [17] [18] [19]. However, this is not true for the MWF cost function. As a result, the trade-off between noise-reduction and preservation of noise binaural cues, provided by a fixed weight parameter (applied into the augmented cost function), may be significantly affected by the absolute power of the received signals (e.g., Lombard effect). Such observation imposes the need for a dynamic weight parameter to maintain the desired trade-off under power variations of the input signals. This approach would alleviate the need for repetitive manual adjustments of the binaural hearing-aid control-parameters to accommodate variations on the acoustic scenario [26].

This work presents a method for the automatic adjustment of MWF-based binaural techniques to ensure the maintenance of the desired noise reduction performance, as well as the preservation of the original perception of the acoustic scenario, under nonstationary conditions. To the best of the author's knowledge, this is the first approach into this area, whose results can be directly applied in a variety of noise reduction methods for binaural hearing aids such as: [16], [17], [19], [27], [28], [29], and [30].

The proposed method is designed for acoustic scenarios comprised of one speech and either an interfering point noise source [16] [17] or diffuse noise field [12]. A theoretical analysis is presented to highlight the mechanisms by which input signal power variations impact the performance of binaural MWF-based methods. Afterwards, we present mathematical properties that should be satisfied by an augmented MWF cost function to present setpoint invariance to input signal power variations. These observations motivate the design of a dynamic weight parameter for achieving robust setpoint invariance. Computational simulations with objective criteria as well as lateralization psychoacoustic experiments with headphones illustrate the performance of the proposed method. The main contributions of this work are: (a) a method for designing the weighting parameter of Wiener-based binaural noise reduction methods for hearing aid applications; (b) a theoretical analysis to elucidate its operating mechanism and support its expected performance; (c) computational simulations with objective measures and real psychoacoustic results to corroborate the effectiveness of the proposed method.

The remainder of this paper is structured as follows: Section II introduces the binaural hearing aid system and the signal model. Section III introduces the problem formulation, while Section IV presents the proposed method. The experimental setup is described in Section V, while results and discussion are, respectively, presented in Sections VI and VII. Finally, concluding remarks are presented in Section VIII. Throughout this text, italic symbols represent scalars. Lowercase and uppercase bold symbols denote, respectively, vectors and matrices. Literals are denoted by non-italic non-bold font. Symbols $\{\cdot\}^{\mathrm{T}}$ and $\{\cdot\}^{\mathrm{H}}$ are, respectively, transpose and Hermitian transpose.

## II. BINAURAL HEARING AID

Fig. 1 illustrates the signal acquisition and processing stages of a binaural hearing aid. The incoming signals are acquired by $M_{\mathrm{L}}$ microphones in the left (L) hearing aid, and by $M_{\mathrm{R}}$ microphones in the right (R) hearing aid. The total number of microphones in the binaural system is $M = M_{\mathrm{L}} + M_{\mathrm{R}}$.

The time-frequency representation of the noisy input signals at a frequency bin index $k$, time frame index $\lambda$, and at all microphones is denoted by the vector $\mathbf{y}(\lambda,k) \in \mathbb{C}^M$, which is given by:

$$\mathbf{y}(\lambda,k) = \mathbf{x}(\lambda,k) + \mathbf{v}(\lambda,k) , \qquad (1)$$

in which $\mathbf{x}(\lambda,k) = \mathbf{h}(\lambda,k)s(\lambda,k)$ is the received speech vector; $\mathbf{h}(\lambda,k) = [\, h_1(\lambda,k)\ h_2(\lambda,k)\ \ldots\ h_M(\lambda,k)\,]^T$ is the speech steering vector, containing the head related transfer functions from the speech source to each microphone; $s(\lambda,k)$ is the clean speech signal at the source; and $\mathbf{v}(\lambda,k)$ is the received additive noise vector. Vector $\mathbf{y}(\lambda,k)$ is available at both hearing aids due to a full duplex communication link, and its entries are given by the left and right input vectors, i.e., $\mathbf{y}(\lambda,k) = [\, \mathbf{y}_L^T(\lambda,k)\ \mathbf{y}_R^T(\lambda,k)\,]^T$, in which $\mathbf{y}_l$ contains the signals received by all $M_l$ microphones in the side $l \in \{L,R\}$.

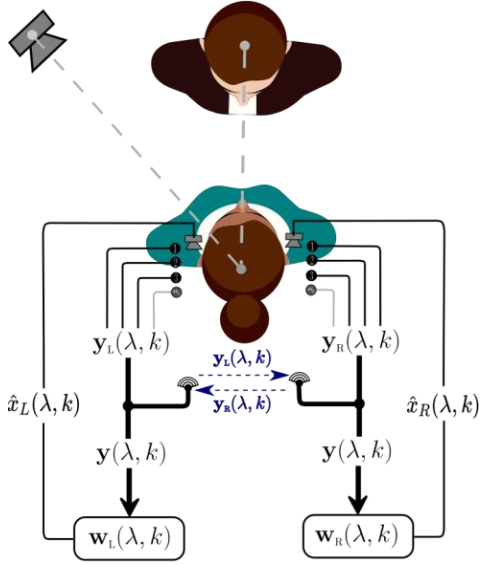

Fig. 1. Binaural hearing-aid system.

In general, the frontal microphone at each side is defined as the reference microphone:

$$y_{l,\text{REF}}(\lambda,k) = x_{l,\text{REF}}(\lambda,k) + v_{l,\text{REF}}(\lambda,k) = \mathbf{q}_l^T \mathbf{y}(\lambda,k) , \qquad (2)$$

in which $\mathbf{q}_l$ is a microphone selection vector with entry equal to 1 in the position of the frontal microphone at side $l$ and zero elsewhere.

The noise reduction problem consists in estimating, for each time-frame and frequency-bin, the desired speech, from the noisy input signals at all microphones. The estimated speech in the left ($\hat{x}_{L,\text{REF}}(\lambda,k)$) and right ($\hat{x}_{R,\text{REF}}(\lambda,k)$) hearing aids, are calculated by the following inner products:

$$\begin{aligned}\hat{x}_{L,\text{REF}}(\lambda,k) &= \mathbf{w}_L^H(\lambda,k)\mathbf{y}(\lambda,k) \\ \hat{x}_{R,\text{REF}}(\lambda,k) &= \mathbf{w}_R^H(\lambda,k)\mathbf{y}(\lambda,k)\end{aligned}, \qquad (3)$$

in which $\mathbf{w}_L(\lambda,k) = [\, w_{L_1}(\lambda,k)\ w_{L_2}(\lambda,k)\ \ldots\ w_{L_M}(\lambda,k)\,]^T$ and $\mathbf{w}_R(\lambda,k) = [\, w_{R_1}(\lambda,k)\ w_{R_2}(\lambda,k)\ \ldots\ w_{R_M}(\lambda,k)\,]^T \in \mathbb{C}^M$ represent the pair of binaural noise reduction filters at the left and right hearing aid, respectively.

*A. Noise Reduction and Binaural Cue Preservation*

The binaural MWF is defined by the minimization of the mean squared error (MSE) between the desired speech at the reference microphones and the processed input signals. Its cost function is defined as:

$$J_{\text{MWF}}(\lambda,k) = E\left\{\left\|\begin{bmatrix}\mathbf{q}_L^T\mathbf{x}(\lambda,k) - \mathbf{w}_L^H(\lambda,k)\mathbf{y}(\lambda,k) \\ \mathbf{q}_R^T\mathbf{x}(\lambda,k) - \mathbf{w}_R^H(\lambda,k)\mathbf{y}(\lambda,k)\end{bmatrix}\right\|^2\right\}, \qquad (4)$$

in which $\|\cdot\|$ is the Euclidean norm, and $E\{\cdot\}$ is the statistical expectation with relation to $x(\lambda,k)$ and $y(\lambda,k)$.

Using (1) in (4), assuming that $\mathbf{w}_L(\lambda,k)$ and $\mathbf{w}_R(\lambda,k)$ are deterministic, and calculating the expectation leads to:

$$\begin{aligned}J_{\text{MWF}}(\lambda,k) = &[\mathbf{q}_L - \mathbf{w}_L(\lambda,k)]^H \mathbf{\Phi}_x(\lambda,k)[\mathbf{q}_L - \mathbf{w}_L(\lambda,k)] \\ &+ \mathbf{w}_L^H(\lambda,k)\mathbf{\Phi}_v(\lambda,k)\mathbf{w}_L(\lambda,k) \\ &+ [\mathbf{q}_R - \mathbf{w}_R(\lambda,k)]^H \mathbf{\Phi}_x(\lambda,k)[\mathbf{q}_R - \mathbf{w}_R(\lambda,k)] \\ &+ \mathbf{w}_R^H(\lambda,k)\mathbf{\Phi}_v(\lambda,k)\mathbf{w}_R(\lambda,k)\end{aligned}, \qquad (5)$$

in which $\mathbf{\Phi}_x(\lambda,k) = E\{\mathbf{x}(\lambda,k)\mathbf{x}^H(\lambda,k)\}$ and $\mathbf{\Phi}_v(\lambda,k) = E\{\mathbf{v}(\lambda,k)\mathbf{v}^H(\lambda,k)\}$ are, respectively, the speech and noise coherence matrices.

The MWF cost function may be complemented by additional penalty terms which aim to preclude solutions that do not preserve the (original) noise spatial-characteristics [12] [16] [17] [18] [19]. This leads to a minimization problem of an *MWF-based augmented cost function* in the form:

$$J_T(\lambda,k) = J_{\text{MWF}}(\lambda,k) + \sum_{i=1}^{I}\alpha_i(\lambda,k)J_i(\lambda,k) , \qquad (6)$$

where $\alpha_i(\lambda,k) \in \mathbb{R}_+$ are weighting (trade-off) parameters which control the balance (minimization effort) among the distinct terms. In general, $J_i(\lambda,k)$ is presented in the following form:

$$J_i(\lambda,k) = E\{|\,BM_{\text{out}}(\lambda,k) - BM_{\text{in}}(\lambda,k)\,|^2\} , \qquad (7)$$

in which $BM \in \{\text{ITD, ILD, ITF, IC}\}$ means "binaural measure"; and subscripts "in" and "out" refer to measures at the input and output of the hearing aids. In general, these measures ($BM_{\text{in}}$ and $BM_{\text{out}}$) are defined in ratio forms, as follows [7] [11]:

$$\begin{aligned}ITD_{\text{in}}(\lambda,k) &= \angle\{\mathbf{q}_L^T\mathbf{v}(\lambda,k)/\mathbf{q}_R^T\mathbf{v}(\lambda,k)\} \\ ITD_{\text{out}}(\lambda,k) &= \angle\{\mathbf{w}_L^H(\lambda,k)\mathbf{v}(\lambda,k)/\mathbf{w}_R^H(\lambda,k)\mathbf{v}(\lambda,k)\}\end{aligned}, \qquad (8)$$

$$\begin{aligned}ILD_{\text{in}}(\lambda,k) &= |\,\mathbf{q}_L^T\mathbf{v}(\lambda,k)/\mathbf{q}_R^T\mathbf{v}(\lambda,k)\,| \\ ILD_{\text{out}}(\lambda,k) &= |\,\mathbf{w}_L^H(\lambda,k)\mathbf{v}(\lambda,k)/\mathbf{w}_R^H(\lambda,k)\mathbf{v}(\lambda,k)\,|\end{aligned}, \qquad (9)$$

$$\begin{aligned}ITF_{\text{in}}(\lambda,k) &= \mathbf{q}_L^T\mathbf{v}(\lambda,k)/\mathbf{q}_R^T\mathbf{v}(\lambda,k) \\ ITF_{\text{out}}(\lambda,k) &= \mathbf{w}_L^H(\lambda,k)\mathbf{v}(\lambda,k)/\mathbf{w}_R^H(\lambda,k)\mathbf{v}(\lambda,k)\end{aligned}, \qquad (10)$$

$$\begin{aligned}IC_{\text{in}}(\lambda,k) &= \mathbf{q}_L^T\mathbf{v}(\lambda,k)\mathbf{v}^H(\lambda,k)\mathbf{q}_L \,/ \\ &\quad |\,\mathbf{q}_L^T\mathbf{v}(\lambda,k)\mathbf{v}^H(\lambda,k)\mathbf{q}_L\,| \\ IC_{\text{out}}(\lambda,k) &= \mathbf{w}_L^H(\lambda,k)\mathbf{v}(\lambda,k)\mathbf{v}^H(\lambda,k)\mathbf{w}_L(\lambda,k) \,/ \\ &\quad |\,\mathbf{w}_L^H(\lambda,k)\mathbf{v}(\lambda,k)\mathbf{v}^H(\lambda,k)\mathbf{w}_L(\lambda,k)\,|\end{aligned}, \qquad (11)$$

in which $\angle\{\cdot\}$ is the angle of its argument. Different solutions may be obtained with distinct approximations for the expectation in (7) [30].

The augmented cost function presented in (6) permits to establish a desired trade-off between noise-reduction/speech-distortion (due to minimization of $J_{MWF}(\lambda,k)$) and binaural noise cue preservation (due to $J_i(\lambda,k)$).

In general, $\alpha_i(\lambda,k)$ are set as a single constant for all time-frames $\lambda$ and frequency bins $k$, but may take into consideration the specific binaural cue frequency prevalence [12] [31].

### III. PROBLEM FORMULATION

An extensive review of the literature indicated that previously proposed augmented MWF techniques employ constant trade-off parameters (i.e., $\alpha_i(\lambda,k) = \beta(k)$) in the cost function presented in (6) [11] [19] [31]. In this way, the performance of the binaural noise reduction method may be influenced by power variations of the received signals [11]. To highlight the influence of such kind of nonstationary behavior, speech and noise signals are represented in a normalized form as:

$$\begin{aligned}\mathbf{v}(\lambda,k) &= g(\lambda,k)\tilde{\mathbf{v}}(\lambda,k)\\ \mathbf{x}(\lambda,k) &= g(\lambda,k)\tilde{\mathbf{x}}_{SNR}(\lambda,k)\end{aligned} \quad (12)$$

in which $g^2(\lambda,k) \in \mathbb{R}_+$ is the mean noise power at the input microphones of the hearing aids, i.e.,

$$g^2(\lambda,k) = E\{\|\mathbf{v}(\lambda,k)\|^2\}, \quad (13)$$

at each bin $k$ and time-frame $\lambda$; $\tilde{\mathbf{v}}(\lambda,k) = \mathbf{v}(\lambda,k)/E\{\|\mathbf{v}(\lambda,k)\|^2\}^{1/2}$ is the normalized noise vector, with unitary variance; $\tilde{\mathbf{x}}_{SNR}(\lambda,k) = SNR_{in}(\lambda,k)^{1/2}\tilde{\mathbf{x}}(\lambda,k)$; $\tilde{\mathbf{x}}(\lambda,k) = \mathbf{x}(\lambda,k)/E\{\|\mathbf{x}(\lambda,k)\|^2\}^{1/2}$ is the normalized speech vector, with unitary variance; and $SNR_{in}(\lambda,k) = E\{\|\mathbf{x}(\lambda,k)\|^2\}/g^2(\lambda,k)$ is the overall input SNR, at each bin $k$ and time-frame $\lambda$, at the microphones.

Using (12) in (1) results in:

$$\begin{aligned}\mathbf{y}(\lambda,k) &= g(\lambda,k) \times [SNR_{in}(\lambda,k)^{1/2} \times \tilde{\mathbf{x}}(\lambda,k) + \tilde{\mathbf{v}}(\lambda,k)]\\ &= g(\lambda,k) \times [\tilde{\mathbf{x}}_{SNR}(\lambda,k) + \tilde{\mathbf{v}}(\lambda,k)]\end{aligned}. \quad (14)$$

Equation (14) models the noisy input-vector by a mutual gain and a weighted sum of two normalized signals. Since $E\{\|\tilde{\mathbf{x}}(\lambda,k)\|^2\} = E\{\|\tilde{\mathbf{v}}(\lambda,k)\|^2\} = 1$, $SNR_{in}(\lambda,k)$ establishes the SNR of the noisy input vector $\mathbf{y}(\lambda,k)$. The parameter $g(\lambda,k)$ is named here as Lombard gain [32], since it is associated to the speech effort ($E\{\|\mathbf{x}(\lambda,k)\|^2\} = SNR_{in}(\lambda,k) \times g^2(\lambda,k)$) that is required to keep the same communication condition (i.e., a constant input SNR). Assuming a fixed SNR, equation (13) establishes a simplified model for representation of the Lombard effect, since it does not take into consideration other spectral modifications [24] [25]. Despite its simplicity, equation (14) constitutes a meaningful way to investigate the impact of input power variations in the performance of MWF-based binaural noise reduction methods. This can be done by considering joint changes of both noise and speech absolute power (constant SNR), as well as modifications of the input SNR.

Now, let us represent $J_{MWF}(\lambda,k)$ in a compact form as

$$J_{MWF}(\lambda,k) = E\{\|[\mathbf{Q}^T - \mathbf{W}^H(\lambda,k)]\mathbf{x}(\lambda,k) + \mathbf{W}^H(\lambda,k)\mathbf{v}(\lambda,k)\|^2\}, \quad (15)$$

in which $\mathbf{Q} = [\mathbf{q}_L \ \mathbf{q}_R]^T$, and $\mathbf{W}(\lambda,k) = [\mathbf{w}_L(\lambda,k) \ \mathbf{w}_R(\lambda,k)]^T \in \mathbb{C}^{2\times M}$. Using (12) in (15) results in:

$$\begin{aligned}J_{MWF}(\lambda,k) &= \\ g^2(\lambda,k) &\times [SNR_{in}(\lambda,k)\|\mathbf{Q} - \mathbf{W}(\lambda,k)\|^2_{\Phi_{\tilde{\mathbf{x}}}^{1/2}(\lambda,k)} \\ &+ \|\mathbf{W}(\lambda,k)\|^2_{\Phi_{\tilde{\mathbf{v}}}^{1/2}(\lambda,k)}]\end{aligned}, \quad (16)$$

in which $\Phi_{\tilde{\mathbf{x}}}(\lambda,k) = E\{\tilde{\mathbf{x}}(\lambda,k)\tilde{\mathbf{x}}^H(\lambda,k)\} = \Phi_{\mathbf{x}}(\lambda,k)/E\{\|\mathbf{x}(\lambda,k)\|^2\}$; $\Phi_{\tilde{\mathbf{v}}}(\lambda,k) = E\{\tilde{\mathbf{v}}(\lambda,k)\tilde{\mathbf{v}}^H(\lambda,k)\} = \Phi_{\mathbf{v}}(\lambda,k)/E\{\|\mathbf{v}(\lambda,k)\|^2\}$; and $\|\mathbf{x}\|_\mathbf{A} = \|\mathbf{Ax}\|$ for $\mathbf{A} \in \mathbb{C}^{M\times M}$.

By considering the ratios that define the binaural cues, presented in (8) to (11), and inspecting equation (7), it becomes easy to verify that the general definition for $J_i$, is not affected by either $g(\lambda,k)$ nor $SNR_{in}(\lambda,k)$, since both the numerator and the denominator in those terms compensate each other.

In the following, the time-frame $\lambda$ and the frequency index $k$ will be omitted in the equations, for clarity reasons, whenever possible.

#### A. Homogeneity Degree

A useful concept to grab theoretical insights of the interplay between the Lombard gain ($g$) and the intrinsic trade-off (speech-distortion and noise reduction versus preservation of the binaural noise cues) associated to the augmented MWF cost function is the *homogeneity degree* [33] of a function, which can be defined as follows:

**Definition 1:** A function $f: \mathbb{C}^M \times \mathbb{C}^M \to \mathbb{R}$ is said to be positively homogeneous of degree $N$ (or $N$-homogeneous), with respect to (w.r.t.) both $\mathbf{a}$, $\mathbf{b} \in \mathbb{C}^M$, for $N \in \mathbb{R}$, if $f(c\mathbf{a},c\mathbf{b}) = c^N f(\mathbf{a},\mathbf{b})$ for all nonzero $c \in \mathbb{R}_+$.

Since the minimum of a function is invariant to positive scaling, the minimum of any $N$-homogeneous function is invariant to joint positive changes in $\mathbf{a}$ and $\mathbf{b}$, i.e., min $f(c\mathbf{a},c\mathbf{b})$ = min $c^N f(\mathbf{a}, \mathbf{b})$ (in which 'min' means 'minimum of'), for $c \in \mathbb{R}_+$. This property can be employed to understand the impact of the Lombard effect (speech and noise joint power variation) over the performance of MWF-based binaural noise reduction methods, as well as to formulate a robust choice for the weighting parameters $\alpha_i(\lambda,k)$ in dynamic environments.

Applying Definition 1 into (16), it is possible to show that $J_{MWF}$ is a 2-homogenous function w.r.t. both $\tilde{\mathbf{x}}_{SNR}$ and $\tilde{\mathbf{v}}$, since:

$$J_{MWF}(g\tilde{\mathbf{x}}_{SNR}, g\tilde{\mathbf{v}}) = g^2 J_{MWF}(\tilde{\mathbf{x}}_{SNR}, \tilde{\mathbf{v}}). \quad (17)$$

On the other hand, it is also possible to show that the penalty terms ($J_i$) using the ratio forms defined by (7) to (11) are 0-homogenous functions w.r.t. $\tilde{\mathbf{v}}$ and $\tilde{\mathbf{x}}_{SNR}$, since

$$J_i(g\tilde{\mathbf{x}}_{SNR}, g\tilde{\mathbf{v}}) = J_i(g\tilde{\mathbf{v}}) = J_i(\tilde{\mathbf{v}}). \quad (18)$$

Using (17) and (18) in (6), and considering a fixed weight parameter $\alpha(\lambda,k) = \beta(k)$, it is easy to verify that the augmented cost function $J_T$ is not homogeneous, since each part has a different degree of homogeneity. As a result, the minimum of $J_T$ is affected by both $g$ and $SNR_{in}$, which means that the desired trade-off between noise reduction and preservation of the binaural cues may change under different acoustic conditions.

### IV. PROPOSED METHOD

In this section, we propose a design method for the weighting parameters $\alpha_i(\lambda,k)$, with the aim of obtaining a robust trade-off between noise-reduction/speech-distortion and binaural-cue preservation, against variations in *g* and *SNR*$_{in}$. As demonstrated in Section III.A, constant weighting parameters $\alpha(\lambda,k) = \beta(k)$ lead to non-homogeneous $J_T$, which is affected by input power variations. Therefore, we propose to employ a dynamic $\alpha$ defined as the multiplication of a constant weight by the noise power (squared Lombard gain), resulting in:

$$\alpha(\lambda, k) = \beta(k) \times g^2(\lambda, k) , \quad (19)$$

in which $\beta(k)$ is a constant (for each *k*) that defines the setpoint associated to the desired trade-off between noise reduction and binaural cue preservation. Using (17), (18), and (19) in (6) results in

$$J_T(g\tilde{\mathbf{x}}_{SNR}, g\tilde{\mathbf{v}}) = g^2 [J_{MWF}(\tilde{\mathbf{x}}_{SNR}, \tilde{\mathbf{v}}) + \beta \sum_{i=1}^{I} J_i(\tilde{\mathbf{v}})]$$
$$= g^2 J_T(\tilde{\mathbf{x}}_{SNR}, \tilde{\mathbf{v}}) \quad (20)$$

Equation (20) shows that the optimal solution ($\mathbf{w}_L$ and $\mathbf{w}_R$) of the resulting augmented MWF-based cost function is invariant to the Lombard gain. This occurs, since in (20) $J_T$ is 2-homogeneous w.r.t. both $\tilde{\mathbf{x}}_{SNR}$ and $\tilde{\mathbf{v}}$. This means that joint variations of speech and noise power should not affect the designed performance of the binaural noise reduction method.

### A. Robustness to $SNR_{in}$ Variations

Although (19) assures robustness against the Lombard effect, the resulting augmented cost function is still non-homogeneous w.r.t. $SNR_{in}$ variations and, as a result, there is no guarantee of setpoint invariance under such condition.

To investigate the impact of $SNR_{in}$ variations in the performance of the proposed method, we employ a heuristic reasoning analysis. By using (16) in (20), the minimization problem to obtain the optimal (left and right) coefficient-vectors can be written as:

$$\min_{\mathbf{W}} J_T = g^2 \min_{\mathbf{W}} \{ \overbrace{SNR_{in} \| \mathbf{Q} - \mathbf{W} \|^2_{\Phi_{\tilde{\mathbf{x}}}^{1/2}} + \| \mathbf{W} \|^2_{\Phi_{\tilde{\mathbf{v}}}^{1/2}}}^{J_{MWF}} \quad (21)$$
$$+ \beta \sum_{i=1}^{I} J_i(\Phi_{\tilde{\mathbf{v}}}) \}$$

It can be noted from (21) that increasing $SNR_{in}$ boosts the significance of $\|\mathbf{Q}-\mathbf{W}\|^2_{\Phi_{\tilde{\mathbf{x}}}^{1/2}}$, which can be interpreted as a weighted similarity measure between $\mathbf{Q}$ and $\mathbf{W}$. Thus, as $SNR_{in}$ increases, the strength of the first term in (21) becomes larger compared to that of the other two, which are not directly influenced by $SNR_{in}$. This leads the solutions to the minimization problem $\mathbf{w}_L$ and $\mathbf{w}_R$ to approximate $\mathbf{q}_L$ and $\mathbf{q}_R$, respectively.

According to [34], which presents a theoretical analysis of the conventional Wiener filter, the amount of noise distortion decreases proportionally to the increase of the input SNR. In fact, for a very large input SNR the optimal solution corresponds to an all-pass filter. It can be shown that this finding, originally presented for the monaural Wiener filter, is also true for the binaural MWF. In this case, increases in the input SNR, lead to decreases in magnitude and phase distortions of the noise at both ears, restoring its original binaural-cues. As a result, the parameter $\beta(k)$ that satisfies the noise binaural cue preservation requirements at a given input SNR, will also satisfy the same requirements at a higher input SNR condition.

### B. Weight Parameter Design

Considering the previous reasoning, devising a parameter design strategy now becomes straightforward. The proposed approach is based on the structure presented in (19), which should provide setpoint invariance to the Lombard gain. Then, considering the worst predicted $SNR_{in}$ condition, the weight parameter $\beta$ is chosen as the smallest value such that the minimum required preservation of the noise binaural-cues is obtained. In this way, higher input SNRs will naturally reinforce the noise binaural-cue preservation. A pseudocode for this procedure is detailed in TABLE I.

TABLE I
PSEUDOCODE FOR THE PARAMETER DESIGN STRATEGY

| **Algorithm 1:** Parameter design strategy |
|---|
| **Input:** $SNR_{worst}$ |
| **Output:** $\beta$ |
| **1.** Generate an acoustic scenario using the worst possible $SNR$ ($SNR_{worst}$) as $SNR_{in}$; For all *k* |
| **2.** Set $\beta(k)=0$ |
| **3.** Compute the solution of the MWF-based method in (21) for $\beta(k)$; |
| **4.** Assess the binaural cue preservation of the noise source at the output; |
| **5. If** the binaural cue preservation is not satisfactory     **Then** increase $\beta(k)$ by a small amount and return to step 3;     **Otherwise** finish; |
| **Return** $\beta(k)$ |

## V. EXPERIMENTAL SETUP

To illustrate the performance of the proposed method we present numerical simulations and psychoacoustic experiments for the MWF with interaural transfer function preservation (MWF-ITF) according to equations (5), (6), (7), (10), and approximations defined in [17]. Comparisons were carried out among unprocessed noise (RAW); noise processed by the MWF method; by the MWF-ITF using a fixed weight parameter; and by the proposed method (MWF-ITF-R). This last one is comprised of equations (13), (19), and the method described in TABLE I.

The main goal of the presented experiments is to illustrate the improved robustness of the proposed method against variations in both Lombard gain and input SNR, better preserving the original perception of the acoustic scenario. The considered acoustic scenario is comprised of a target speech source and a single point noise source. Sub-indexes were added to refer to speech (S), noise (N), and left (L) and right (R) ears.

### A. Acoustic Scenario

The acoustic scenario was simulated using head-related impulse responses (HRIRs) obtained from an anechoic chamber using a manikin Bruel & Kjær type 4128-C wearing a pair of behind-the-ear hearing aids [35]. Each hearing aid has 3 microphones, resulting in $M = M_L + M_R = 6$ microphones. The speech source was assumed to be located at an azimuth $\theta_S = 0°$ and distance $r_S = 0.8$ m from the manikin, while the noise source was at $\theta_N = -60°$ (left side) and $r_N = 3.0$ m.

*B. Speech and Noise Signals*

A set of 32 speech audio files from the repeated Harvard database [36] was selected. The sentence "the fruit peel was cut in thick slices" was uttered by the same female speaker. A set of 32 non-overlapping segments of ICRA noise type I [37], which is a speech-like noise with spectral and temporal properties similar to those of the human voice, was employed for contamination. Fig. 2(a) and Fig. 2(b) illustrate the speech and noise spectrograms, respectively. Fig. 3 details the estimated power spectrum density (PSD) of both speech and noise. The sampling frequency was $f_s = 16$ kHz.

The noisy signals at the hearing aid microphones were created by convolving speech and noise, respectively, with $HRIR_S$ and $HRIR_N$, and then summing the weighted signals to obtain a desired input SNR.

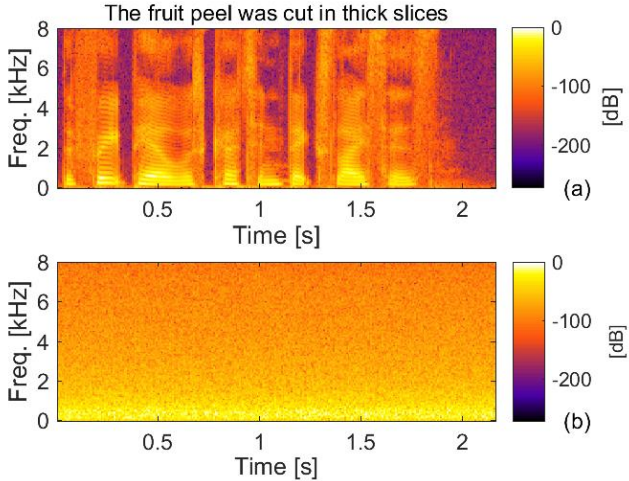

Fig. 2. Spectrograms of: (a) speech sentence: "the fruit peel was cut in thick slices"; and (b) one segment of the ICRA noise type I.

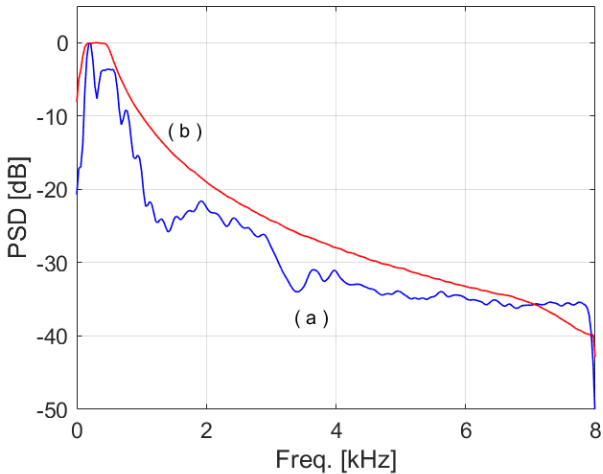

Fig. 3. Estimated power spectral density (PSD): (a) speech (blue); and (b) ICRA type I noise (red).

The time-frequency representation of the input signals was obtained using the short-time Fourier transform (STFT) with a frame size of 128 samples, corresponding to a time-frame of 8 ms. A squared-root Hanning window was applied. The discrete Fourier transform employed $K = 256$ bins, zero padding, and 50% overlap.

The noise coherence matrices ($\mathbf{\Phi}_v(\lambda,k)$) were calculated directly from the available noise epochs. This was performed to reduce reinforcement of the binaural cues of the processed signals due to estimation errors [38]. In real applications, this coherence matrix can be estimated as in [39]. The noisy signal coherence matrices ($\mathbf{\Phi}_y(\lambda,k)$) were obtained from the noisy-signal as in real applications. The speech coherence matrix was approximated by $\mathbf{\Phi}_x(\lambda,k) = \mathbf{\Phi}_y(\lambda,k) - \mathbf{\Phi}_v(\lambda,k)$.

The noisy-input, speech, and noise signals were processed, in the frequency domain, by the filter coefficients obtained by applying the Broyden-Fletcher-Goldfarb-Shanno quasi-Newton optimization method [40] [41] to minimize the MWF, MWF-ITF, and MWF-ITF-R cost functions. The time-domain synthesized output signals were generated by weighting adjacent frames by the synthesis window [42].

*C. Objective Criteria*

Simulation results were assessed according to six objective criteria. The left and right global input-output SNR differences, which measures the global SNR variation at each ear, are calculated as:

$$\Delta SNR_l = 10\log_{10}\left(\sum_{k=1}^{K}\frac{\mathbf{w}_l^H(k)\mathbf{\Phi}_x(k)\mathbf{w}_l^H(k)}{\mathbf{w}_l^H(k)\mathbf{\Phi}_v(k)\mathbf{w}_l^H(k)}\right) \\ -10\log_{10}\left(\sum_{k=1}^{K}\frac{\mathbf{q}_l^T\mathbf{\Phi}_x(k)\mathbf{q}_l}{\mathbf{q}_l^T\mathbf{\Phi}_v(k)\mathbf{q}_l}\right), \quad (22)$$

in which $\mathbf{\Phi}_x(k)$ and $\mathbf{\Phi}_v(k)$ assume stationary signals. The ILD and ITD differences [30] for both speech and noise are defined as:

$$\Delta ITD_e = \left(\sum_{k=1}^{k_s}\angle\frac{\mathbf{w}_L^H(k)\mathbf{\Phi}_e(k)\mathbf{w}_L(k)}{\mathbf{w}_R^H(k)\mathbf{\Phi}_e(k)\mathbf{w}_R(k)}\right) \\ -\left(\sum_{k=1}^{k_s}\angle\frac{\mathbf{q}_L^T\mathbf{\Phi}_e(k)\mathbf{q}_L}{\mathbf{q}_R^T\mathbf{\Phi}_e(k)\mathbf{q}_R}\right), \\ \Delta ILD_e = 10\log_{10}\left(\sum_{k=k_s+1}^{K/2+1}\frac{\mathbf{w}_L^H(k)\mathbf{\Phi}_e(k)\mathbf{w}_L(k)}{\mathbf{w}_R^H(k)\mathbf{\Phi}_e(k)\mathbf{w}_R(k)}\right) \\ -10\log_{10}\left(\sum_{k=k_s+1}^{K/2+1}\frac{\mathbf{q}_L^T\mathbf{\Phi}_e(k)\mathbf{q}_L}{\mathbf{q}_R^T\mathbf{\Phi}_e(k)\mathbf{q}_R}\right), \quad (23)$$

in which $e \in \{S, N\}$, $K$ is the number of bins, and $k_s = \lfloor 1500K/f_s \rfloor$ is the highest integer smaller than $1500K/f_s$.

The averaged squared Lombard gain ($\overline{g}^2$) is estimated according to

$$\overline{g}^2 = \frac{1}{\Lambda}\sum_{\lambda=1}^{\Lambda}\sum_{k=1}^{K}\hat{g}^2(\lambda,k) = \frac{1}{\Lambda}\sum_{\lambda=1}^{\Lambda}\sum_{k=1}^{K}\hat{\sigma}_v^2(\lambda,k), \quad (24)$$

in which $\Lambda$ is the number of available time-frames; and the estimated $SNR_{in}$ ($\overline{SNR}_{in}$) is calculated as:

$$\overline{SNR}_{in} = \frac{\sum_{\lambda=1}^{\Lambda}\sum_{k=1}^{K}\hat{\sigma}_x^2(\lambda,k)}{\sum_{\lambda=1}^{\Lambda}\sum_{k=1}^{K}\hat{\sigma}_v^2(\lambda,k)}, \qquad (25)$$

*D. Psychoacoustic Experiments*

Psychoacoustic experiments were conducted to assess the performance of volunteers in a lateralization task. A headphone Sennheiser HD 202 was connected to a laptop, by using a Realtek® high-definition audio onboard sound card, under Windows 10 operational system. Noise-only signals processed by the assessed methods were employed to quantify the perceived azimuth of arrival. A total of 16 volunteers participated in the experiment, which were divided in two groups. In the first group, employed for Lombard gain analysis, there were 2 females and 6 males, resulting in an average age of 25.6 years with a standard deviation of 5 years. In the second group, employed for analyzing SNR variations, three females and five males performed the experiment. The average age of this group was 23.2 years and a standard deviation of 4 years. The experiments were approved by the Ethics Committee in Human Research, under certificate 90899518.7.0000.0121 CEP-UFSC. All volunteers read and signed the written informed consent form.

After an initial period necessary for adjusting the volume to comfortable levels and providing instructions, the experiment procedure was divided into three stages: (a) learning: in which the volunteers listened to audios associated to visual information about their respective azimuths, (from −90° to 90°, spaced by 15°, including the edge points); (b) training: in which volunteers were instructed to associate seven audios with a set of seven azimuths: {−90°, −60°, −30°, 0°, 30°, 60°, 90°} – during this stage volunteers were allowed to listen to the audios and to change their choices as many times as they wanted before finishing the experiment; and (c) testing: in which the volunteers were instructed to listen to a set of randomly selected audios and to blindly assign them to an azimuth. The volunteers were allowed to listen to the audios as many times as they wanted before assigning.

Unprocessed noise (RAW), as well as, MWF, MWF-ITF, MWF-ITF-R processed noise were obtained according to the acoustic scenario described in Section V.A, for four different noise runs, and either $0 \text{ dB} \leq \bar{g}^2 \leq 30 \text{ dB}$ or $-5 \text{ dB} \leq SNR_{in} \leq 25 \text{ dB}$ in steps of 5 dB [20] [43]. Additionally, five unprocessed noise-only audios related to the following azimuths {−90°, −30°, 0°, 30° 90°} were also employed, totalizing 117 audios for each volunteer in each experiment.

VI. RESULTS

In this section, the performance of the proposed method was assessed by both objective quality metrics and psychoacoustic experiments, considering either Lombard-gain or SNR variations. The acoustic scenario was created according to Sections V.A and V.B. TABLE II shows the input signal quality measures. The proposed method assumed the following arbitrary design requirements: $SNR_{worst} = -5$ dB [20] [43]; and a binaural cue preservation requirement for the noise source defined by $\Delta ILD_N < 2$ dB and $\Delta ITD_N < 2$ ms, which resulted in $\beta(k) = 1$ for all $k$. All volunteers were able to assign each noise signal to their correct azimuth in stage 2 (training) of the psychoacoustic experiment, showing adequate individual lateralization capacity.

TABLE II
INPUT QUALITY MEASURES

| Measure | Average values |
|---|---|
| $PESQ_L$ | 1.05 |
| $PESQ_R$ | 1.07 |
| $SNR_L$ [dB] | −6.71 |
| $SNR_R$ [dB] | −2.60 |
| $SNR_{in}$ [dB] | −5 |

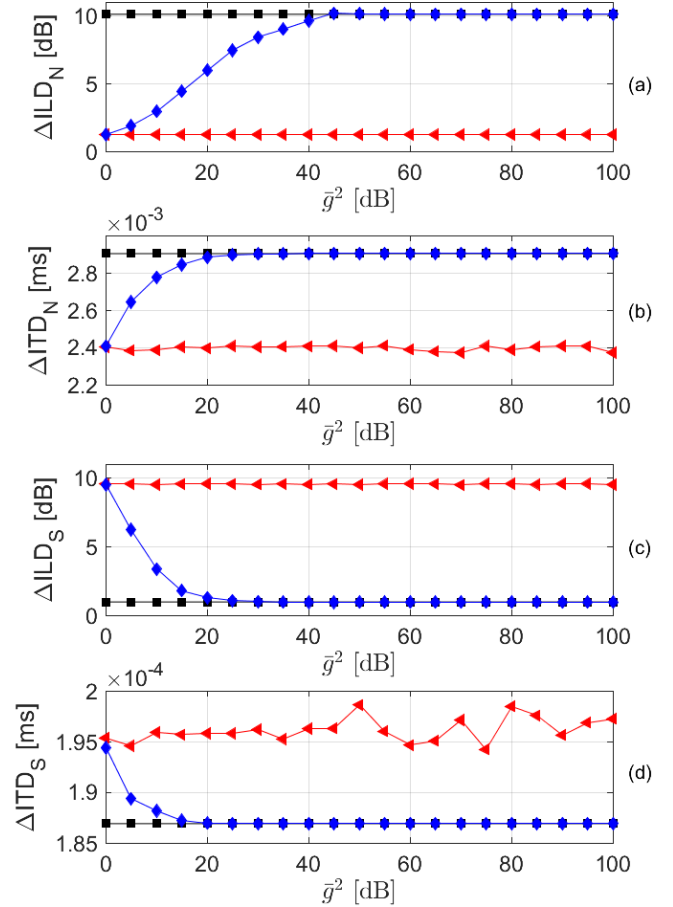

Fig. 4. Averaged binaural-cue errors with respect to the averaged squared Lombard gain ($\bar{g}^2$) in deciBels (dB) for: (i) MWF (black square), (ii) MWF-ITF (blue diamond), and (iii) MWF-ITF-R (red triangle). (a) $\Delta ILD_N$ [dB]; (b) $\Delta ITD_N$ [ms]; (c) $\Delta ILD_S$ [dB]; (d) $\Delta ITD_S$ [ms].

*A. Robustness to the Lombard Gain: Numerical Simulations*

Fig. 4 shows speech and noise $\Delta ILD$ and $\Delta ITD$ values for MWF, MWF-ITF, and MWF-ITF-R processed signals, as a function of the average squared Lombard gain (see (24)). It is clearly noted that the performance of the conventional MWF-ITF is significantly affected by variations in the Lombard gain.

Fig. 4 also shows that both MWF and MWF-ITF-R are approximately invariant to changes in $\bar{g}^2$. Considering $\Delta ILD_N$ and $\Delta ITD_N$, respectively shown in Fig. 4(a) and Fig. 4(b), the MWF presents the highest noise distortion, while the MWF-ITF-R presents the lowest. On the other hand, for $\Delta ILD_S$ and $\Delta ITD_S$, respectively shown in Fig. 4(c) and Fig. 4(d), the MWF results in the lowest speech distortion, and the MWF-ITF-R in the highest. A similar behavior is verified for $\Delta SNR_L$ and the $\Delta SNR_R$ presented, respectively, in Fig. 5(a) and Fig. 5(b).

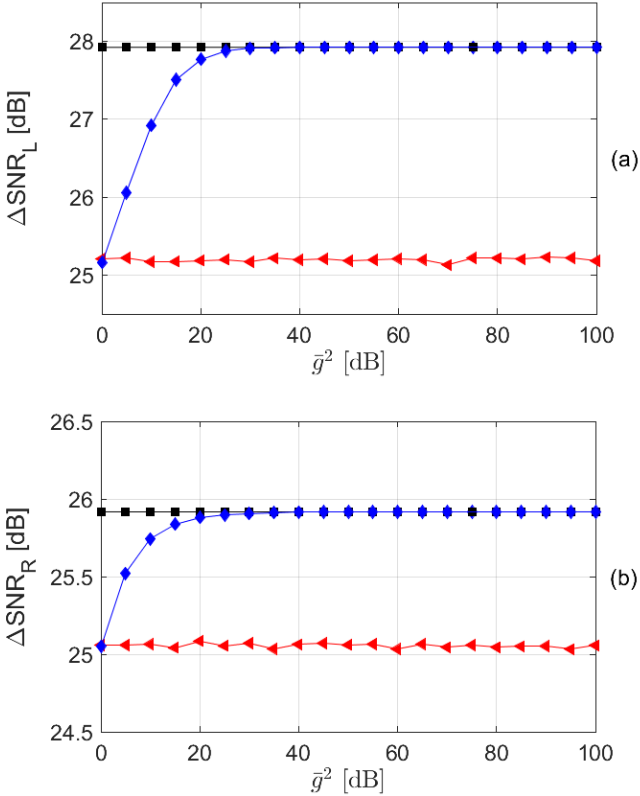

Fig. 5. Averaged input-output SNR differences with respect to the average squared Lombard gain ($\bar{g}^2$) in deciBels (dB) for: (i) MWF (black square), (ii) MWF-ITF (blue diamond), and (iii) MWF-ITF-R (red triangle). (a) $\Delta SNR_L$ [dB]; and (b) $\Delta SNR_R$ [dB].

### B. Robustness to the Lombard Gain: Psychoacoustic Experiments

Fig. 6(a) and Fig. 6(b) present, respectively, the average and median perceived azimuth of arrival for each noise reduction method and unprocessed noise, obtained in the psychoacoustic experiments with volunteers. It can be noted that the average/median perceived azimuth obtained from signals processed by the MWF-ITF is significantly affected by the Lombard Gain.

Fig. 6(c) presents box-and-whisker diagrams including all Lombard gains. The average perceived azimuth for the unprocessed (RAW) noise is −71° (median of −75°). In this sense, we verify that volunteers overestimate the true azimuth of the noise source localized at −60° (left). As expected for signals processed by the MWF, the perceived noise azimuth is changed towards the speech source at azimuth 0° (average of 3° and median of 0°). The MWF-ITF presents an average perceived azimuth of −25° (median of −19°), indicating a displacement towards the speech azimuth. This is especially true for large Lombard gains, when its cost function is largely unbalanced. The proposed MWF-ITF-R results in accurate estimates of the original noise azimuth with an average of −54° (median of −56°).

Considering the average response of the volunteers for each audio, and that each $\bar{g}^2$ indicates a different condition in which the azimuth was measured, a one-way repeated measures ANOVA test was applied to analyze the three methods with relation to the perceived azimuth. Firstly, the Shapiro-Wilk (SW) test was applied to the data of each method to verify the hypothesis of gaussianity. Its null hypothesis is defined as "$H_0$: the distributions are Gaussian" at the level of significance $p > 0.05$. $H_0$ was accepted for all groups. The Mauchly's test was used to analyze the sphericity assumption of the ANOVA test. The null hypothesis of this test is "$H_0$: differences between all possible pairs are equal" and the level of significance was $p > 0.05$. The sphericity assumption was rejected, and the Greenhouse-Geisser (GG) correction was applied. As no extreme outliers were identified, the one-way repeated ANOVA, with the null hypothesis defined as "$H_0$: all distributions are the same, at the level of significance $p > 0.05$" was applied. As a result, the null hypothesis was rejected ($F(2,68) = 66.5$, $p < 0.0001$, $\eta^2 = 0.56$). Then, the Dunn-Bonferroni post-hoc test was applied to verify which pairs of distributions were different from each other. The null hypothesis was "$H_0$: the pairs of groups have the same distributions, at a level of significance $p > 0.05$". Finally, we concluded that all pairwise differences, between methods, were statistically different ($p \leq 0.05$).

Speech azimuth was undoubtedly perceived at 0° for all methods, not being further analyzed.

### C. Robustness to SNR Variations: Numerical Simulations

A preliminary computational simulation was performed for corroborating the theoretical hypothesis presented in Section IV.A, with relation to the robustness of equation (21) against $SNR_{in}$ variations.

Fig. 7 shows the ratio between the average (for all bins) of $J_{MWF}$ and $\beta \cdot \Sigma^I_{i=1} J_i(\mathbf{\Phi}_{\tilde{v}})$, for $I = 1$, $BM \equiv ITF$ and optimal $\mathbf{W}$, in (21), i.e.,

$$\eta = \left(\sum_{k=1}^{K} J_{MWF}(k)\right) \Big/ \left(\beta \sum_{k=1}^{K} J_1(k)\right), \quad (26)$$

as a function of $SNR_{in}$ for $\beta \in \{0, g^{-2}, 1\}$, respectively corresponding to the MWF, MWF-ITF, and the MWF-ITF-R methods. From Fig. 7, it is possible to verify that for $\beta \in \{0, g^{-2}\}$ (MWF and MWF-ITF) the ratio in (21) substantially decreases with increasing $SNR_{in}$. For $SNR_{in} > 0$ dB, $\eta$ decays logarithmically with $SNR_{in}$. On the other hand, for $\beta = 1$ (MWF-ITF-R) $\eta$ is approximately constant for all range of $SNR_{in}$. From these observations it can be inferred that ensuring preservation of the binaural noise cues for a given SNR condition, assures that they will also be preserved for higher SNRs.

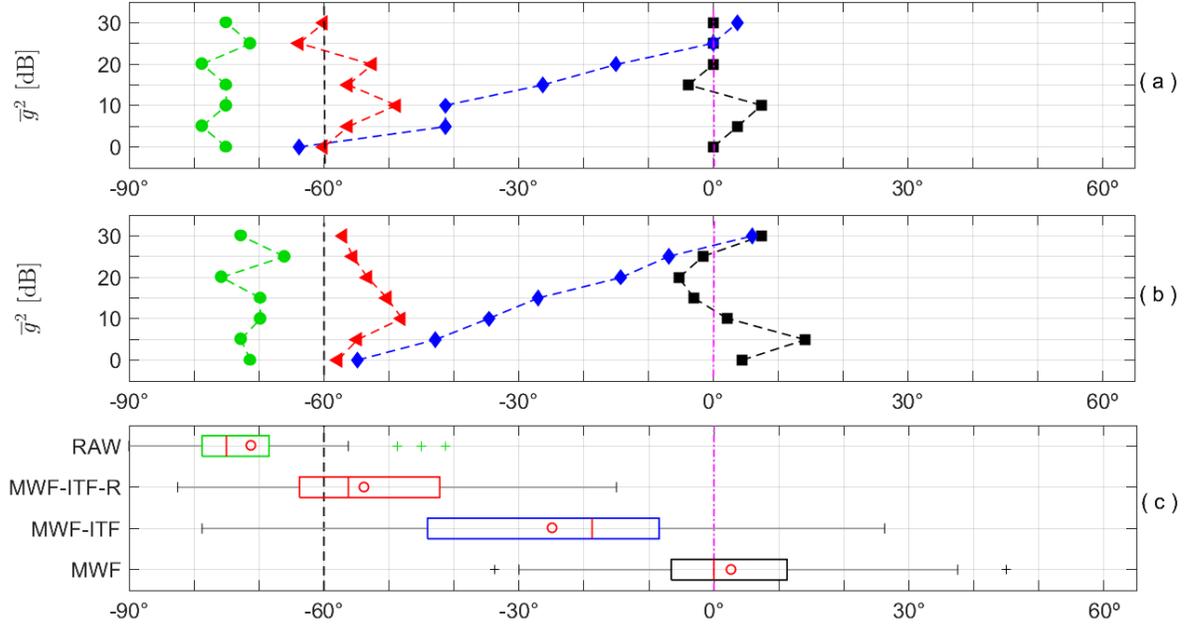

Fig. 6. Psychoacoustic experiments for different average squared Lombard gains: RAW (unprocessed) (green circle), MWF (black square), MWF-ITF (blue diamond), MWF-ITF-R (red triangle), true noise azimuth (black dashed line), and true speech azimuth (pink dotted-dashed line). (a) Average and (b) median perceived azimuths. (c) Box-and-whisker diagrams for all perceived azimuths grouped by technique (circles represent the average values).

Following, Fig. 8 shows speech and noise $\mathit{\Delta ILD}$ and $\mathit{\Delta ITD}$ as a function of $\overline{SNR}_{in}$. Both MWF-ITF and MWF-ITF-R methods have the same performance at $\overline{SNR}_{in} = -5$ dB. Speech binaural cues for both MWF-ITF and MWF-ITF-R are similar, and not very distinct from MWF results, indicating that the azimuth of the target source is preserved.

It can be observed that both $\mathit{\Delta ILD}_N$ and $\mathit{\Delta ITD}_N$ results are affected by the $\overline{SNR}_{in}$. The binaural cues of the noise processed by the proposed method are mostly affected in the $-50$ dB $< \overline{SNR}_{in} < -5$ dB range, while for the conventional MWF-ITF this range is $-5$ dB $< \overline{SNR}_{in} < 60$ dB.

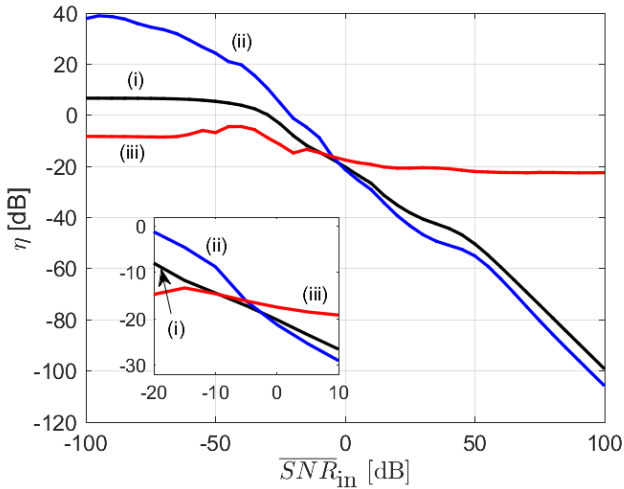

Fig. 7. Simulation results for equation (26): (i) $\beta = 0$ (MWF) (black); (ii) $\beta = g^{-2}$ (MWF-ITF) (blue); and (iii) $\beta = 1$ (MWF-ITF-R) (red).

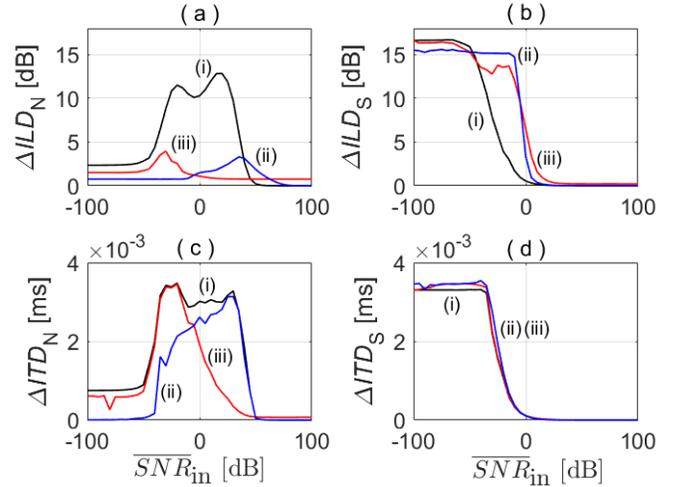

Fig. 8. Influence of the $\overline{SNR}_{in}$ in the binaural noise cue preservation: (i) MWF (black); (ii) MWF-ITF (blue); and (iii) MWF-ITF-R (red). (a) $\mathit{\Delta ILD}_N$, (b) $\mathit{\Delta ILD}_S$, (c) $\mathit{\Delta ITD}_N$, and (d) $\mathit{\Delta ITD}_S$.

### D. Robustness to SNR Variations: Psychoacoustic Experiments

Fig. 9(a) and Fig. 9(b) present, respectively, the average and median azimuth perceived by volunteers for each noise reduction method and unprocessed noise as a function of $\overline{SNR}_{in}$. It can be noted that the azimuth perception obtained from both MWF and MWF-ITF processed signals are more affected by the $\overline{SNR}_{in}$ than the MWF-ITF-R. Fig. 9(c) presents box-and-whisker diagrams of all perceived azimuths grouped by processing method. The average (and median) perceived azimuth for the unprocessed noise (RAW) is $-68°$. In this sense, volunteers overestimate the real azimuth of the noise source as also verified in the psychoacoustic experiments in Section V.B.

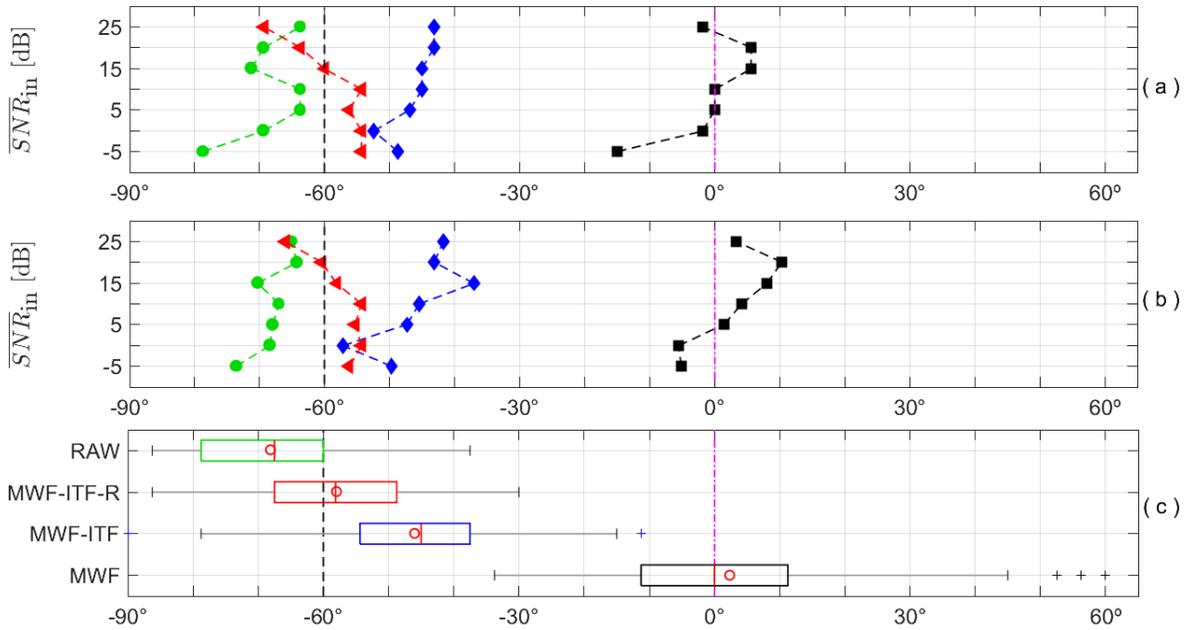

Fig. 9. Psychoacoustic experiments for different $\overline{SNR}_{in}$: RAW (unprocessed) (green circle), MWF (black square), MWF-ITF (blue diamond), and MWF-ITF-R (red triangle), true noise azimuth (black dashed line), and true speech azimuth (pink dotted-dashed line). (a) Average and (b) median perceived azimuths. (c) Box-and-whisker diagrams for all perceived azimuths grouped by technique (circles represent the average values).

As expected, the perceived azimuths provided by the MWF are severely displaced toward to the speech source at azimuth 0° (average of 0° and median of 2°). The MWF-ITF resulted a perceived average azimuth at −46° (median of −45°), while MWF-ITF-R provided an average (and median) of −58°.

Unprocessed and processed noise by the MWF, MWF-ITF and MWF-ITF-R methods were analyzed with relation to their perceived azimuth performance, constituting four groups. No extreme outliers were identified in any group. Gaussianity was accepted, but sphericity assumption was rejected. In this way, the Greenhouse-Geisser transformation was applied. The one-way repeated ANOVA test was applied, in which the null hypothesis is "$H_0$: all distributions are the same", at the level of significance $p > 0.05$. The null hypothesis was rejected. Using the Dunn-Bonferroni post-hoc test, all pair comparisons rejected the null hypothesis. Finally, we concluded that all pairwise differences, between methods, were statistically different ($p \leq 0.05$).

## VII. Discussion

The proposed dynamic weighting parameter and its designing method aim to provide a robust operating setpoint, for a desired noise reduction effort and preservation of the original perceived azimuth for both speech and noise acoustic sources, under nonstationary conditions.

In Section III, the influence of both the Lombard effect and the SNR condition into augmented MWF-based cost functions was theoretically presented.

Numerical simulations with objective criteria indicate that the proposed method presented in Section IV is robust to both the Lombard effect (Fig. 4 and Fig. 5) and to input SNR variations (Fig. 8).

Fig. 4 and Fig. 5 show that the performance of the MWF-ITF is significantly affected by the Lombard gain. It can be noted that increasing the Lombard gain increases both $\Delta SNR_L$ and $\Delta SNR_R$, as well as preservation of the speech binaural cues. However, both $\Delta ILD_N$ and $\Delta ITD_N$ indicate a progressive distortion of the noise binaural cues. For $\overline{g}^2 > 45$ dB the MWF-ITF performance is equivalent to that of the conventional MWF, and the spatial preservation of the noise source is lost. As a result, the desired trade-off between noise-reduction and preservation of the binaural cues is no longer achieved. On the other hand, the MWF-ITF-R performance is approximately constant keeping the same binaural-cue preservation independently of the whole range of assessed Lombard gains.

Fig. 6 shows psychoacoustic experiments with volunteers, corroborating the observations presented in Fig. 4(a) and Fig. 4(b). It can be noted that changes in the Lombard gain significantly modify the perception of the noise azimuth in signals processed by the MWF-ITF. The perceived azimuth ranges vary in average from −64° to 4° (with an average of −25°) for $0$ dB $\leq \overline{g}^2 \leq 30$ dB. The MWF-ITF-R provides the most accurate noise azimuth perception (varying from −64° to −49°, with an average of −54°), while the unprocessed-noise azimuth is overestimated by the volunteers (varying from −79° to −71°, resulting in an average of −71°).

Fig. 7 corroborates the theoretical hypothesis that the proposed weighting parameter provides robustness to input SNR variations, showing bounded limits to the balance of both parts (noise reduction and spatial preservation) of the MWF-ITF technique.

Fig. 8 indicates that both the MWF-ITF and the MWF-ITF-R present approximately the same binaural-cue variation along the analyzed range of $\overline{SNR}_{in}$. However, the largest amounts for each method occur in different ranges. The MWF-ITF shows large binaural cue distortions for both $\Delta ILD_N$ and $\Delta ITD_N$ in the 5 dB $< \overline{SNR}_{in} <$ 55 dB range, while the MWF-ITF-R is only significantly affected in $-40$ dB $< \overline{SNR}_{in} < -15$ dB. This observation agrees with the theoretical assumption that the robustness of the proposed method increases with $\overline{SNR}_{in} \to \infty$. It is important to note that according to [20], the most common listening situations in which adults with hearing losses are exposed are in the 2 dB $< \overline{SNR}_{in} <$ 14 dB range. In this way, the proposed method provides better spatial preservation as compared to the conventional MWF-ITF under practical conditions.

Finally, Fig. 9 also corroborates both theoretical assumptions and numerical simulations with regard to robustness to input SNR variations. In the same way as in Fig. 6, the unprocessed-noise azimuth is overestimated by the volunteers (varying, in average, from $-79°$ to $-64°$, for $-5$ dB $\leq \overline{SNR}_{in} \leq 30$ dB), resulting in a global average/median of $-68°$. The MWF results in average azimuths from $-15°$ to $6°$, and global average of $2°$ (median of $0°$, which is the speech azimuth), while the MWF-ITF results in a range from $-53°$ to $-43°$, and global average of $-46°$ (median of $-45°$). The MWF-ITF-R method provides the most accurate azimuth perception (varying from $-69°$ to $-54°$, with global average/median of $-58°$).

The variation of the perceived azimuth in the psychoacoustic experiments is expected since its average error increases for more lateral angles [7] [14].

## VIII. Conclusion

This work proposed a dynamic weighting parameter for MWF-based binaural noise-reduction methods in hearing aid applications. It aims to provide robust preservation of the spatial localization cues under speech and noise power variations. A designing method was presented to guarantee that the desired operational setpoint, which establishes a tradeoff between noise reduction and binaural-cue preservation, is preserved. An application example using the MWF-ITF noise reduction method (which can be easily expanded to the ILD, ITD, and IC cases) demonstrated that the proposed technique is robust to both the Lombard Effect and input SNR variations. Simulation results with objective criteria as well as psychoacoustic experiments with volunteers corroborate the theoretical arguments, indicating that the proposed method keeps the desired trade-off between noise-reduction/speech-distortion and preservation of the speech and noise binaural-cues, even under challenging conditions.